\documentclass{article}

\pdfoutput=1 

\usepackage{PRIMEarxiv}

\usepackage[utf8]{inputenc} 
\usepackage[T1]{fontenc}    
\usepackage{hyperref}       
\usepackage{url}            
\usepackage{booktabs}       
\usepackage{amsfonts}       
\usepackage{nicefrac}       
\usepackage{microtype}      
\usepackage{lipsum}
\usepackage{tabularx}
\usepackage{fancyhdr}       
\usepackage{graphicx}       
\graphicspath{{media/}}     

\title{A plano-convex thick-lens velocity map imaging apparatus for direct, high resolution 3D momentum measurements of photoelectrons with ion time-of-flight coincidence}

\author{
  Michael Davino$^{1}$, Edward McManus$^{1}$, Nora G. Helming$^{1}$, Chuan Cheng$^{2}$, \\
  \textbf{G\"{o}nen\c{c} Mo\v{g}ol$^{2}$, Zhanna Rodnova$^{1}$, Geoffrey Harrison$^{1}$, Kevin Watson$^{1}$,} \\
  \textbf{Thomas Weinacht$^{2}$, George N. Gibson$^{1}$, Tobias Saule$^{1}$, Carlos Trallero-Herrero$^{1,}$*} \\
  $^{1}$University of Connecticut, Department of Physics, Storrs, CT, 06269 \\
  $^{2}$Stony Brook University, Department of Physics and Astronomy, Stony Brook, 11794, USA \\
  \texttt{*carlos.trallero@uconn.edu} \\
}

\begin{document}
\maketitle

\begin{abstract}
Since its inception, velocity map imaging (VMI) has been a powerful tool for measuring the 2D momentum distribution of photoelectrons generated by strong laser fields. There has been continued interest in expanding it into 3D measurements either through reconstructive or direct methods. Recently much work has been devoted to the latter of these, particularly by relating the electron time-of-flight (TOF) to the third momentum component. The technical challenge here is having timing resolution sufficient to resolve structure in the narrow (< 10 ns) electron TOF spread. Here we build upon work in the fields of VMI lens design and 3D VMI measurement by using a plano-convex thick-lens (PCTL) VMI in conjunction with an event-driven camera (TPX3CAM) providing TOF information for high resolution 3D electron momentum measurements. We perform simulations to show that, with the addition of a mesh electrode to the thick-lens VMI geometry, a plano-convex electrostatic field is formed which extends the detectable electron cutoff energy range while retaining high resolution. Further, the thick-lens also extends the electron TOF range which allows for better resolution of the momentum along this axis. We experimentally demonstrate these capabilities by examining above-threshold ionization (ATI) in Xenon where the apparatus is shown to collect electrons of energy up to $\sim$7 eV with a TOF spread of $\sim$30 ns, both of which are improvements on previous work by factors of $\sim$1.4 and $\sim$3.75 respectively. Finally, the PCTL-VMI is equipped with a coincident ion TOF spectrometer which is shown to effectively extract unique 3D momentum distributions for different ionic species within a gas mixture. These techniques have potential to lend themselves to more advanced measurements, particularly involving systems where the electron momentum distributions possess non-trivial symmetries and require high resolution.
\end{abstract}

\section{Introduction}

Velocity map imaging (VMI) \cite{eppink1997velocity} is a widely used and extremely versatile technique for measuring a large range of ion or electron energies with high resolution\cite{ding2021composite,zhang2016aplanatic,kling2014thick,kregel2017multi}. The VMI technique utilizes a series of cylindrically symmetric electrodes to create an electrostatic field which serves as lens to map charged particles of a given transverse momentum (or energy) to a specific position on the surface of a detector. Over time a variety of lensing geometries have been developed to increase the relative energy resolution and range to > 1 keV \cite{ding2021composite,kling2014thick} and < 2\% \cite{kling2014thick,zhang2016aplanatic} respectively. One such development pushing previous limits is the thick-lens VMI by Kling et al. \cite{kling2014thick}. This design was shown to extend the VMI detectable energy range while maintaining high resolution by using 11 VMI electrodes rather than the standard 3. Additionally, new applications of mesh electrodes have become common \cite{kauczok2009three,ablikim2019coincidence,zhang2014method,ding2021composite}. These electrodes allow for equipotential surfaces that span the electrostatic lens and give fine control of the field while being transparent enough to limit particle loss.

Though these changes have enhanced VMI capabilities, the net result of VMI largely remained a 2D projection of the 3D momentum distribution. Despite this traditional limit of VMI, there has been a constant and continued interest in attaining the full 3D momentum distribution for use in studying various phenomena such as ionization \cite{pengel2017electron,maurer2012molecular} and photodissociation \cite{dinu2002application,chichinin2002three}. Other 3D methods such as cold-target recoil-ion-momentum spectroscopy (COLTRIMS) \cite{ullrich2003recoil,ullrich1997recoil,dorner2000cold} have proven effective to this end, but can be prohibitively expensive and difficult to implement relative to VMI.

3D VMI momentum measurements have been achieved either through reconstruction techniques or, more recently, direct measurement. Of the two methods, direct measurement is generally considered the ideal in most circumstances as reconstruction often comes with considerable drawbacks. The most common reconstruction technique utilizes the inverse Abel transform \cite{shepp1974fourier,strickland1991reconstruction} which requires cylindrical symmetry and is thus unsuitable for experiments using elliptically polarized light. Alternatively, tomographic imaging methods have been implemented \cite{smeenk2009momentum} which recover 3D information by rotating the momentum distribution (via the exciting laser field polarization) and reconstructing from multiple 2D projections. This has the obvious drawback of requiring multiple datasets to be taken to recover a single 3D distribution which in not practical in some situations, e.g., pump-probe experiments. Further, both reconstruction methods are liable to introduce noise and potentially artifacts into the reconstructed distribution.

Direct measurement of the 3D distribution can be performed by using the particle’s time-of-flight (TOF) which encodes its momentum along the axis of the electrostatic lens. DC slicing \cite{gebhardt2001slice} is a technique by which the VMI detector is gated on a range of TOFs by a fast voltage switch as to image only a ‘slice’ of the 3D momentum distribution. Scanning the gate then allows for the full 3D distribution to be measured. Unfortunately, not only does this method require multiple datasets to be taken to produce a single distribution, but it also effectively discards the majority of data by measuring only the small fraction of particles that fall within a given slice.

The most straightforward, though technically challenging, method for measuring the 3D distribution is by coincident measurement of each particle’s position and TOF. In this method the momentum components (p$_{x}$,p$_{y}$,p$_{z}$) are measured by (x,y,t) respectively, where x and y are the position of the particle on the surface of a detector and t is its TOF. This type of measurement has attracted significant interest in recent years despite the high temporal resolution (generally < 1 ns) needed to measure TOF, particularly with respect to electrons. Basnayake et al. performed 3D momentum measurements for electrons by correlating images with the output of a photomultiplier tube resulting in a temporal resolution of $\sim$30 ps \cite{basnayake2022three}. Cheng et al. accomplished similar measurements by coupling the signal from the phosphor anode of a microchannel plate (MCP) detector into an event-driven camera, TPX3CAM (TPX3) \cite{Amsterdam_instruments_2022}, which has two built-in time-to-digital converters (TDCs) \cite{cheng20223d}. The TPX3 TDCs (and overall detection) are limited to a TOF resolution of 260 ps, but with the tradeoff that the scheme is easily implementable on almost any VMI and requires no synchronization between devices. Further, Cheng et al. performed 3D momentum measurements for electrons and ions in coincidence using a single detector with a fast-switching voltage supply.

Although traditional (2D) VMI can measure electrons with transverse kinetic energy on the order of 1 keV, doing so requires increasing the gradient of the electrostatic field which reduces the TOF range of the particles; in turn this necessitates unrealizable temporal resolutions to achieve 3D measurement. This effectively means that measurements of this kind are limited to low energy while still requiring temporal resolution on the order of 100 ps. For reference, Cheng et al. run their VMI at voltages which limit it to the collection of electrons with kinetic energy less than $\sim$5 eV, but the electron TOF spread is still only $\sim$8 ns.

Here we expand on the work of both Cheng et al.\cite{cheng20223d} and Kling et al.\cite{kling2014thick} by using a TPX3 camera in conjunction with a coincident plano-convex thick-lens (PCTL) VMI and ion TOF spectrometer for measurement of 3D electron momentum distributions. By simply introducing a grounded mesh electrode at the far end of a thick-lens VMI, the electrostatic field forms a plano-convex lens which extends the detectable electron cutoff energy range while maintaining high spatial energy resolution. Additionally, the thick-lens serves to increase the electron TOF range, thus requiring less temporal resolution in electronics. The net result is the collection of electrons with transverse energies up to $\sim$7 eV while extending the electron TOF spread to $\sim$30 ns. Further, the coincident ion TOF measurement allows for significant noise reduction and is shown to be effective in extracting the unique 3D electron momentum distributions for multiple ion species measured simultaneously. These techniques for high resolution 3D momentum information have potential to lend themselves to more advanced measurements such as 3D holography , photoionization with intricate exciting fields (e.g. elliptical, two-color), and other systems for which electron momentum distributions are not cylindrically symmetric.

\begin{figure*}[ht!]
\includegraphics[width=\linewidth]{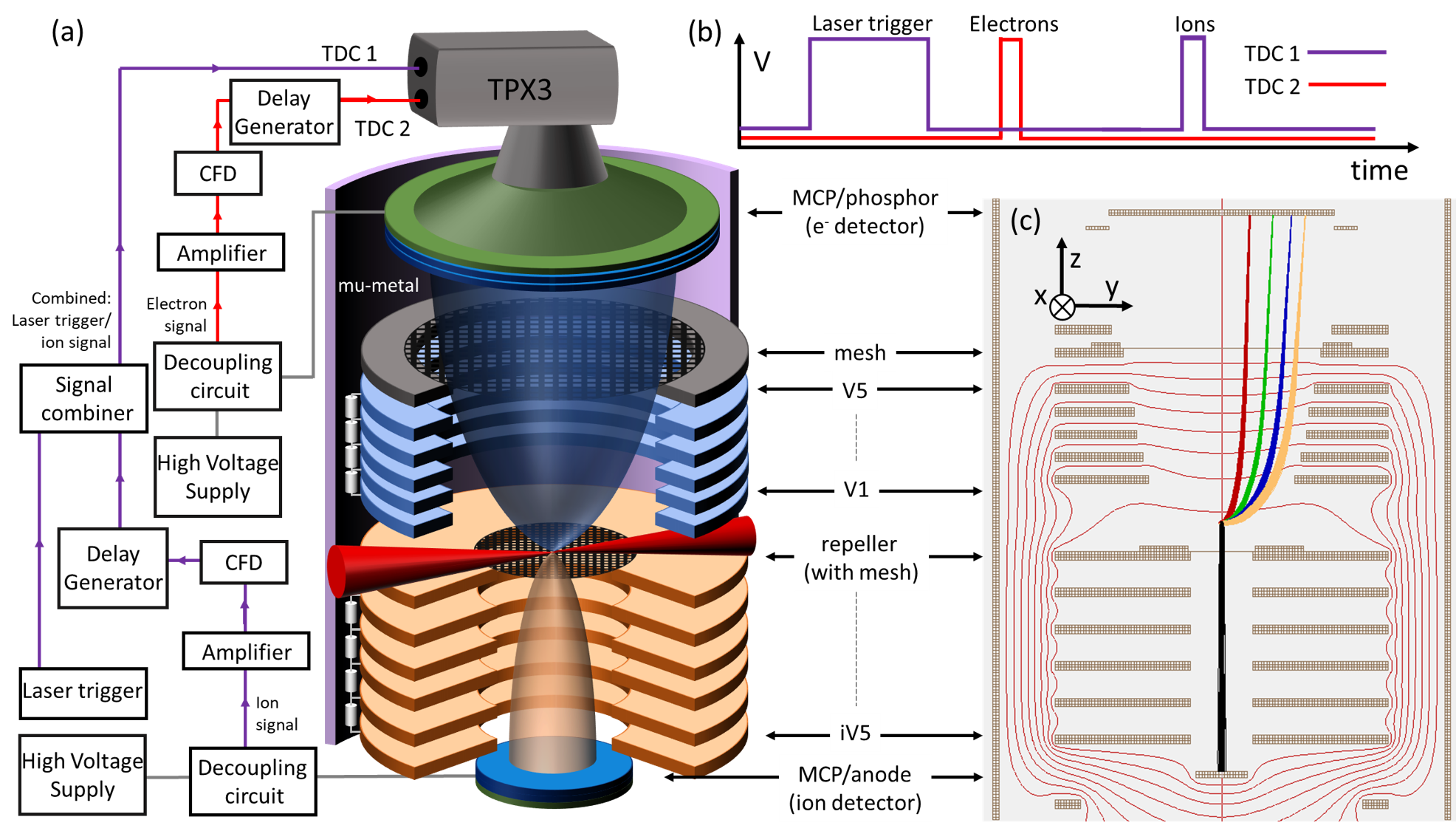}
\caption{\label{fig:apparatus}(a) is a depiction of the coincident PCTL-VMI and ion TOF spectrometer and its signal processing electronics. Electron-side electrodes are shown in blue and ion-side electrodes in orange. The repeller electrode is equipped with a mesh, as is an additional, grounded plate on the far end of the electron-side (grey). The lens assembly and detectors are encased in a mu-metal cylinder. Electrons are detected by an MCP stack with a phosphor screen anode and imaged by a TPX3 camera with two built in TDC inputs. Ions are detected by an MCP stack and stainless-steel anode. Both detectors are supplied high voltage through a decoupling circuit which allows the signal pulse from each detected particle to be separated and subsequently amplified. The amplified signals are each passed through a constant fraction discriminator (CFD) and a delay generator, the result of which is a positive TTL pulse. The output ion TTL pulses are subsequently combined with a reference laser trigger (also TTL) and the combined signal is recorded by TDC 1 of the TPX3 camera; the electron pulses are sent directly into TDC 2 of the TPX3 after the delay generator. (b) is a temporal depiction of the two TDC signals shown in a plot of voltage as a function of time. For each laser pulse, the laser trigger arrives first in TDC 1, followed by the electron signal in TDC 2, and then the ion signal which is also in TDC 1. (c) is a schematic of the VMI cross section in SIMION. In red are equipotential lines which show the plano-convex lens of the electron side. Electron trajectories for four different energies are shown in red, green, blue and yellow, and ion trajectories are shown in black.}
\end{figure*}

\section{Apparatus}

The general setup of the coincident PCTL-VMI and ion TOF spectrometer is depicted in Figure \ref{fig:apparatus}. Note the axes shown in Figure \ref{fig:apparatus} (c) are used throughout this manuscript, particularly, $\hat{z}$ is the axis of the electrostatic lens and $\hat{x}$ is the axis of laser propagation.

The electrostatic lens is formed by eleven stainless steel electrodes, five on the electron side (V1 through V5) and six on the ion side (repeller, iV1 through iV5). The position and potential of each electrode is detailed in Table \ref{tab:electrodes}. Both the electron and ion side electrodes are connected in a chain by 100 M$\Omega$ resistors, but the two sides are isolated from one another. In doing so, only four voltages need to be applied to the PCTL-VMI lens, one at either end of each electrode chain (V1, V5, repeller, and iV5), and the equal resistance resistors ensure a uniform voltage step between every two adjacent electrodes in each chain. The electron side geometry is based on a previous design \cite{kling2014thick}.

\begin{table}[b!]
\centering
\begin{tabularx}{\linewidth}{p{75pt}||p{75pt}|p{75pt}}
\hline
\textbf{Electrode} & \textbf{Position (mm)} & \textbf{Potential (V)} \\
\hline
\hline

Mesh & 106 & 0 \\
\hline

V5 & 94 & -37.3 \\
\hline

V4 & 89 & -49.7 \\
\hline

V3 & 84 & -62.1 \\
\hline

V2 & 79 & -74.6 \\
\hline

V1 & 74 & -87 \\
\hline

Repeller & 50 & -100 \\
\hline

iV1 & 40 & -100.2 \\
\hline

iV2 & 30 & -100.4 \\
\hline

iV3 & 20 & -100.6 \\
\hline

iV4 & 10 & -100.7\\
\hline

iV5 & 0 & -100.9\\
\end{tabularx}
\caption{\label{tab:electrodes}Geometry of the coincident PCTL-VMI and ion TOF spectrometer. Blue and orange shaded cells comprise the electron and ion side electrodes respectively. Electrodes for each side are connected by a series of equal resistance resistors to achieve a uniform step in potential across each two adjacent electrodes when voltages are applied at either end of the chain (V1 and V5, or repeller and iV5).}
\end{table}

In addition to the electrodes there is a grounded, copper mesh disk on the far end of the electron side. The mesh is 12 mm from the V5 electrode and 71 mm in diameter; it is commercially available from Precision eforming with 3.15 lines per mm and 80\% transmission efficiency. The grid is crucial to the formation of the plano-convex electrostatic lens on the electron side as it serves to be the ‘plano’ of the lens by providing a uniform voltage spanning the transverse profile of the lens. This is in contrast to typical annular electrodes which result in a convex field. The plano-convex field is clearly visible in Figure \ref{fig:apparatus} (c), a SIMION schematic of the VMI cross section in which the red lines indicate equipotential surfaces. Each of the five electrodes on the electron side are annular and, moving away from the interaction region, with each electrode the field lines become flatter up to the surface of the grid. The repeller is equipped with the same mesh except 23 mm in diameter; such a mesh on the repeller is not uncommon for a double sided VMI (or VMI-TOF). Note that when a mesh is referred to throughout the rest of this manuscript it is referring to the electron-side mesh, not the repeller mesh.

The electron detector is a stack of two 80 mm diameter microchannel plates (MCPs) in chevron configuration equipped with a P47 phosphor anode. The front face of the electron detector is grounded so, given that the mesh is also grounded, there is a 45 mm field free drift region between the exit of the plano-convex lens and the electron detector. The ion detector is a stack of two 25 mm MCPs (also in chevron configuration) equipped with a stainless-steel anode.

The lens electrodes, mesh, and detectors are encased in a mu-metal cylinder which shields the charged particles from any external magnetic field. The entire assembly is mounted in a high vacuum chamber with a baseline pressure of 2.5 x 10$^{-8}$ Torr.

External to the vacuum, the electron MCP and phosphor stack is imaged by a TPX3 camera which is equipped with two time-to-digital converters (TDC 1 and 2), both with a temporal resolution of 260 ps. Additionally, the TPX3 also natively records the time at which each pixel is activated (time of arrival, TOA) with a resolution of 1.6 ns. In our detection scheme TDC 1 is used to measure both the TOF for ions and a laser trigger, while TDC 2 is used for electron TOF. In this regard, the TPX3 camera represents a convenient platform for multiple laser-event synchronization. Pixel TOA is used to distinguish laser correlated electron events from noise by removing pixel events outside of a 1 $\mu$s window for each laser pulse. Further, because a pixel TOA is recorded for every pixel activation, a clustering algorithm is implemented to identify multiple pixel events as belonging to the same electron by evaluating their proximity in space and time. Clustering serves to effectively increase the spatial resolution of the camera \cite{bromberger2022shot}. Here we make use of an open-source algorithm, DBSCAN \cite{ester1996density}. Because we are grouping pixel events into clusters, we will associate each cluster with a cluster TOA that is the mean TOA, weighted by the duration of pixel activation (time-over-threshold, TOT), for the pixel events comprising that cluster. In practice, TOA is recorded by the TPX3 as an absolute time, however, throughout the rest of this manuscript we will refer to TOA as it is processed and referenced to a specific laser pulse. In other words, TOA is now the time difference between a laser timing reference signal and the recorded (absolute) TOA. This effectively makes the cluster TOA a measure of the electron TOF, but the 1.6 ns resolution is insufficient to resolve structure in p$_{z}$, so its use is limited to noise reduction.

A diagram of the electronic schemes for electron and ion detection is shown in Figure \ref{fig:apparatus} (a). Both the electron and ion detector anodes are supplied voltage through simple decoupling circuits which allow for single particles to be detected electronically. Signal pulses from the decoupling circuits are amplified and subsequently processed by constant fraction discriminators (CFDs, Ortec models 9327 and 583) with temporal jitter < 20 ps and < 75 ps for the electrons and ions respectively. Because the TPX3 TDCs call for a positive TTL pulse, the output of each CFD triggers a delay generator (SRS models DG535 for ions and DG645 for electrons) to produce such a pulse. We use two delay generators to avoid dead-time issues in the processing of the signals. All together the temporal resolution of the electronics configurations for both electrons and ions is limited by the resolution of the TDCs themselves at 260 ps.

In an event-driven experiment it is necessary to label each event; we achieve this by recording each laser pulse non-ambiguously using the TDC 1 channel of the TPX3. For this purpose, the ion TTL signal is first combined with a laser trigger TTL through a DC pass splitter/combiner (Mini-Circuits, model ZX10R-2-183-S+). This combined ion and laser trigger signal is critical as the current TPX3 model has only two TDCs, but to attain the 3D momentum distribution we require both the ion and electron signals, as well as the laser trigger to serve as reference for the electron and ion TOF. The laser trigger TTL pulse arrives well before the ion signal ($\sim$260 $\mu$s) meaning there is no concern of overlap, and has a significantly longer pulse width relative to the ion TTL (1 $\mu$s versus 50 ns) so that the two are easily distinguishable. The time sequence of the three signals as recorded by the two TDCs is shown in Figure \ref{fig:apparatus} (b).

\section{Results}

\subsection{Simulations}

To optimize the operating voltages and determine the energy range and resolution capabilities of the PCTL-VMI geometry shown in Figure \ref{fig:apparatus} (c), electron trajectory simulations were performed using SIMION \cite{dahl1990simion} (version 8.1). A key feature of the PCTL-VMI apparatus presented here is the capability to directly measure the 3D momentum distribution of electrons which requires the range and resolution of the PCTL-VMI to be assessed in two parts. First, the momentum of the electron transverse to the axis of the electrostatic lens (xy-plane) is related to the position of the electron on the MCP/phosphor detector; this is the standard operation mode for a 2D VMI. The corresponding (spatial) energy resolution, $\Delta$E/E, is proportional to the range of radii (distance from center of the detector, R) for electrons of a given energy. Specifically, because a VMI exhibits a linear relationship between the electron momentum and R, we can approximate $\Delta$E/E $\simeq$ 2$\Delta$R/R as is standard\cite{kling2014thick}. Here we take $\Delta$R to be the full width at half maximum (FWHM) of the electron radial distribution.

\begin{figure}[b!]
\centering
\includegraphics[width=80mm]{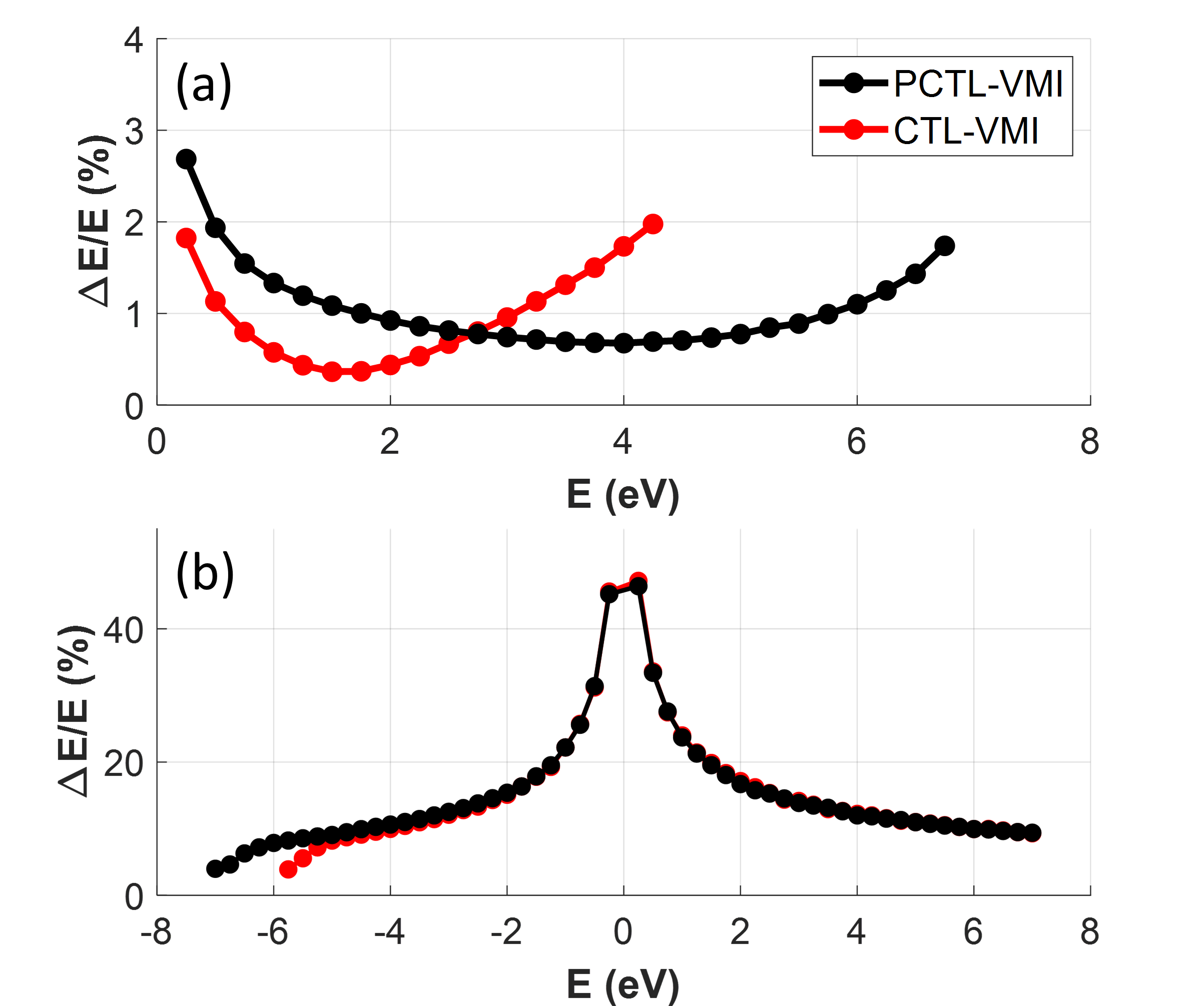}
\caption{\label{fig:SIMION} Predicted energy resolution by SIMION simulations for spatially (a) and temporally (b) resolved electron energies. Note that negative energies in (b) correspond to electrons with velocity in -$\hat{z}$. Simulations were run with the thick-lens VMI for both plano-convex (PCTL-VMI) and conventional convex (CTL-VMI) electrostatic lens configurations, for a fixed -100 V applied to the repeller. The source volumes was taken as a 1 mm radius, 2 mm length cylinder and 5000 particles were flown for each electron energy.}
\end{figure}

Additionally, the component of momentum along the axis of the detector, p${_z}$, is measured by the electron TOF. Assessing this (temporal) $\Delta$E/E is less straightforward as there is not a linear relationship between the TOF and momentum. Further, for electrons emitted with momentum $\pm$$p_{z}$, there is a lack of symmetry up vs down in the electrostatic field. Thus, to evaluate the temporal energy resolution we require an additional simulation to serve as a reference and reliably relate a given electron TOF with its initial momentum. This reference simulation is performed with electrons originating from a single, ‘ideal’, point with energies from 0 eV to 7 eV in steps of 0.001 eV and initial velocity in both $\pm$$\hat{z}$. The reference simulation is presented later in the text as a comparison to experimental results and can be seen in Figure \ref{fig:calibration} (d). Using this simulation to correlate electron TOF to momentum $\Delta$E/E is now evaluated in a manner similar to before: $\Delta$E is the energy range corresponding to the FWHM of the TOF distribution for a given kinetic energy.

The simulated spatial and temporal energy resolutions of the PCTL-VMI are presented in Figure \ref{fig:SIMION} (a) and (b) respectively. The electron source volume was taken to be a cylinder of 1 mm radius and 2 mm length with its axis along $\hat{x}$. Electrons have initial momentum corresponding to kinetic energies from 0 eV to 7.5 eV in steps of 0.25 eV; this range is particularly relevant as it corresponds to low (up to fifth) order ATI from 800 nm light. The initial electron momenta are along $\hat{y}$ to determine the spatial energy resolution (Figure \ref{fig:SIMION} (a)) or $\pm$$\hat{z}$ for the temporal energy resolution (Figure \ref{fig:SIMION} (b)). Note that in Figure \ref{fig:SIMION} (b) positive and negative kinetic energies correspond to initial momenta in the respective $\pm$$\hat{z}$ direction.

Each plot of $\Delta$E/E for the PCTL-VMI shown in Figure \ref{fig:SIMION} is accompanied by simulation results for the same VMI geometry except with a conventional (convex) thick-lens (labelled CTL-VMI). This conversion from plano-convex to convex is achieved simply by removing the mesh and reoptimizing the electrode voltages while keeping the repeller voltage fixed at -100 V. Note that for both configurations the electrode voltages are optimized with the restriction that they are connected by a chain of equal resistance resistors, that is, there is a uniform voltage drop between adjacent electrodes. Thus, for a fixed repeller voltage, the optimization of the electron-side lens is a two-parameter problem consisting of the voltages applied on the ends of the electrode stack (electrodes V1 and V5). With optimized electrode voltages, it is shown in Figure \ref{fig:SIMION} (a) that both the PCTL-VMI and CTL-VMI configurations have spatial energy resolution better than or comparable to the resolution of most modern VMI designs over their respective energy ranges \cite{kling2014thick,zhang2014method,zhang2016aplanatic,ding2021composite}.

Although electrons were simulated with energies up to 7.5 eV, Figure \ref{fig:SIMION} (a) shows that, for initial velocities in $\hat{y}$ and the repeller at -100 V, the PCTL-VMI collects electrons up to a cutoff of 6.75 eV as compared to just 4.25 eV for the CTL-VMI. Further, while the CTL-VMI has about twice the resolution of the PCTL-VMI for electron energies between 0.5 eV and 2 eV, the PCTL-VMI configuration retains a high energy resolution ($\Delta$E/E < 1\%) over a range more than twice that of its convex counterpart. The energy resolution curve of the CTL-VMI is tunable by the voltage on V1 such that the $\Delta$E/E minimum can be shifted between 1.5 eV and 3.5 eV, but the cutoff energy cannot be extended without significantly compromising the resolution. In fact, tuning the voltage for better resolution at the higher end of this energy range further limits the cutoff energy by about 0.75 eV ($\sim$18\%). The PCTL-VMI configuration exhibits a similar dependence on the V1 electrode but because it already has a high energy resolution over much of its collection range, optimizing the $\Delta$E/E curve for specific energies comes at the cost of resolution over the rest of the range. In general, the PCTL-VMI is shown more valuable for reliable spatial energy resolution over a large range of electron energies, whereas the CTL-VMI may be preferred in instances where extremely high resolution is required over a narrow range of energies. With respect to 3D momentum measurements, measurement of p$_{x}$ and p$_{y}$ is already extremely limited by the low voltages required to maintain a resolvable TOF spread. In this case the PCTL-VMI design offers a means to extend the detectable energy range without increasing voltages, thus retaining a broad TOF range.

\begin{figure*}[b!]
\includegraphics[width=\linewidth]{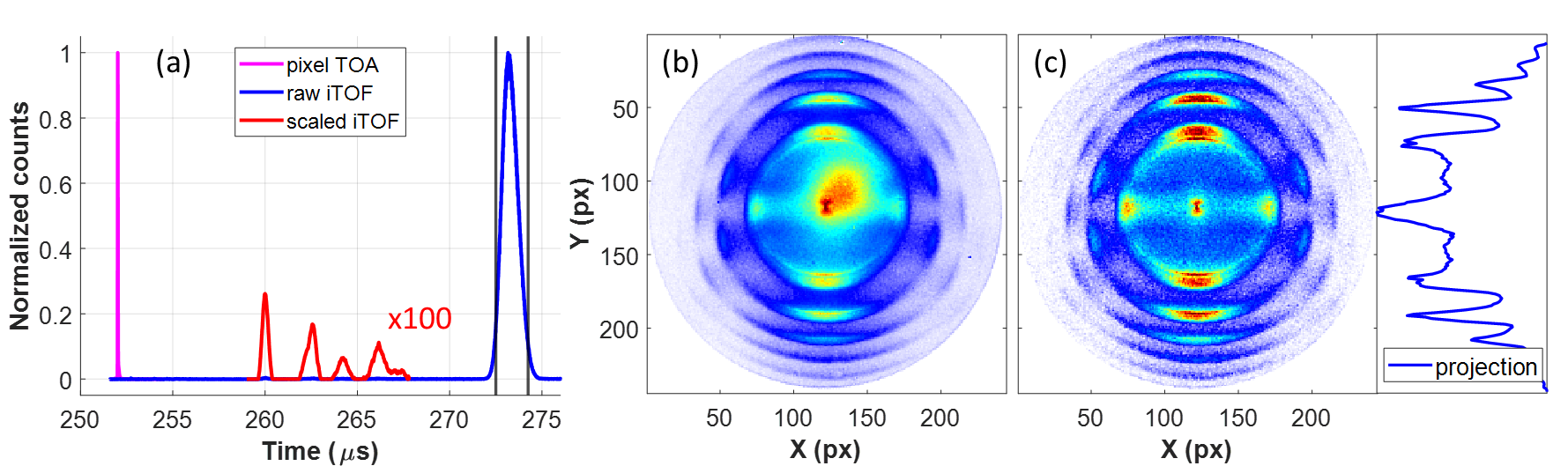}
\caption{\label{fig:raw} Ionization of Xe gas with linearly polarized light oriented in $\hat{y}$ for $\sim$600k laser pulses. (a) is a histogram of cluster TOA and ion TOF (iTOF); also overlain is a portion of the TOF spectrum scaled by a factor of 100 to visualize other ion species which are present. (b) is a clustered VMI image with no gating used, i.e., every pixel activation included, while (c) is a clustered VMI image with gating on both cluster TOA (1 $\mu$s gate, centered on the pixel TOA peak in (a)) and ion TOF (1.75 $\mu$s, vertical black lines in (a)). (c) also shows the projection of this image onto the polarization axis, $\hat{y}$.}
\end{figure*}

Figure \ref{fig:SIMION} (b) shows the temporal energy resolution for the two configurations. Here the PCTL and CTL are almost indistinguishable as the electron TOF is most dependent on the potential difference between the repeller and V1 rather than the lensing itself. The temporal energy resolution is shown to be less than its spatial resolution, but it compares well to the resolution of some current, conventional VMI designs (e.g. this design by Garcia et al. \cite{garcia2013delicious}). There are two other points to be made regarding Figure \ref{fig:SIMION} (b); first, although the simulated electrons only have velocity along the axis of the lens ($\hat{z}$), it is shown that the CTL-VMI cannot collect electrons with velocities corresponding to more than $\sim$5.8 eV in -$\hat{z}$ (E < -5.8 eV in Figure \ref{fig:SIMION} (b)). This is because, given enough velocity along -$\hat{z}$, an electron can escape the region between the repeller and V1 into the (largely) field free ion side of the apparatus and will not have a trajectory that ends at the electron detector. To this end, the PCTL does have a slight advantage over the CTL because its optimized V1 voltage is slightly less negative than in the CTL configuration. On the other hand, although trajectories for electrons with initial velocity in +$\hat{z}$ will always reach the electron detector and are shown to have reasonable energy resolution, this does not mean that in practice we can resolve energies in this regime. The issue is that although $\Delta$E/E appears reasonable, the differences in the electron TOF become so small for similar energies that the TPX3 TDC does not have the temporal resolution to distinguish them. We have found (experimentally) for our setup we are able to resolve energetic structures for electrons up to $\sim$3.5 eV with velocities along +$\hat{z}$ and $\sim$7.5 eV for velocities along -$\hat{z}$ (see Figure \ref{fig:calibration} (b) and (d)).

\subsection{Experiment}

\begin{figure*}[b!]
\centering
\includegraphics[width=150mm]{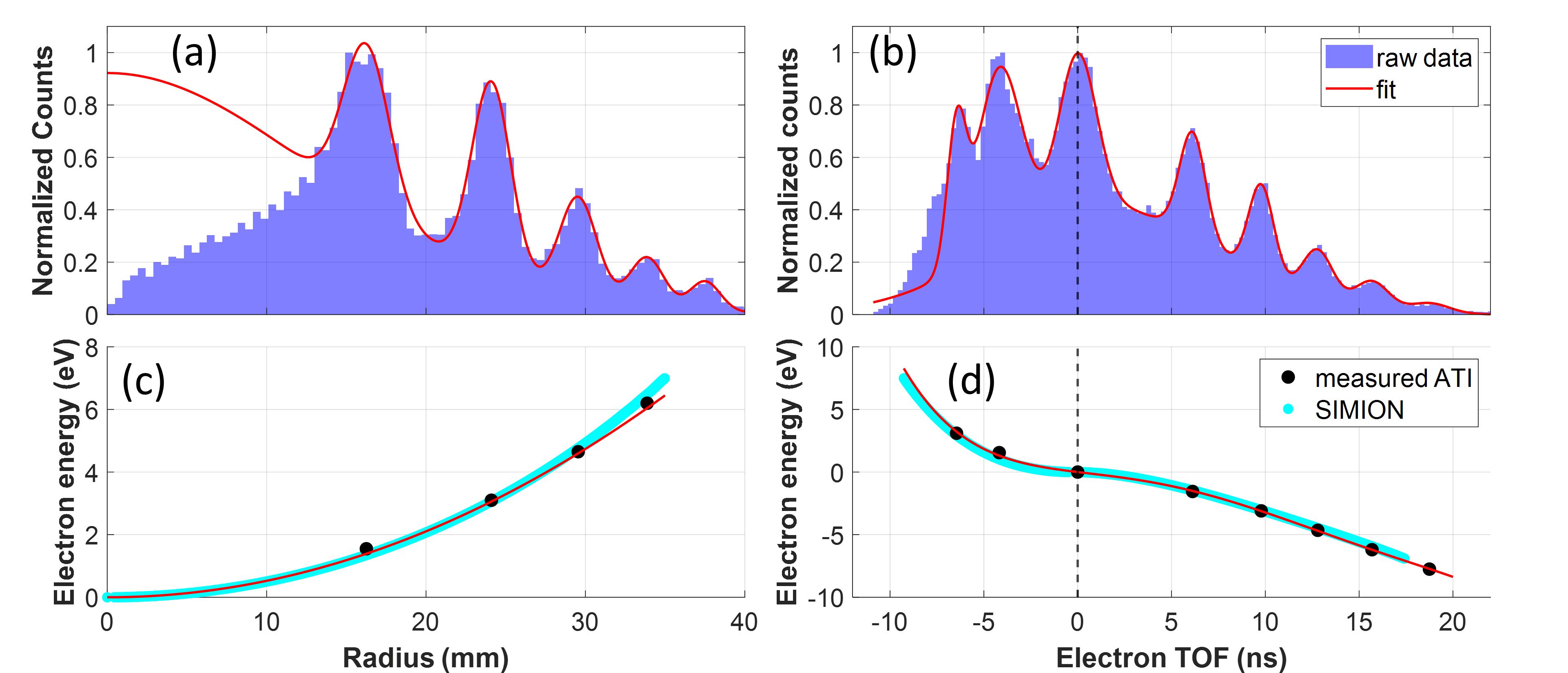}
\caption{\label{fig:calibration} Spatial and temporal energy calibration of the PCTL-VMI. (a) is a normalized histogram of the radius (distance from center of mass) for cluster events corresponding to Xe singly ionized by linearly polarized laser pulses oriented in $\hat{y}$ (image shown in Figure \ref{fig:raw} (c)). It is fit by the sum of multiple Gaussian functions to find the position of ATI peaks. The ATI peaks are plotted against their energy and fit to a quadratic then compared to SIMION results in (c). Note that though five ATI peaks are fit in (a), the fifth is close to the edge of the detector and thus omitted from the calibration in (c). Similarly, (b) is a histogram of electron TOF (eTOF) for Xe singly ionized by linearly polarized laser pulses but now oriented in $\hat{z}$. This is also fit to a sum of Gaussian functions and subsequently the ATI peaks are fit to a fifth order polynomial and compared to SIMION results in (d). Note that the x-axis in (b) and (d) has been shifted such that a TOF of zero corresponds to zero kinetic energy (denoted by a dashed vertical line in (b) and (d)). Also note that negative energies correspond to initial velocities in -$\hat{z}$.}
\end{figure*}

To demonstrate the 3D measurement capabilities of PCTL-VMI, above-threshold ionization (ATI) was observed in Xenon (Xe). For these experiments Xe gas was injected into the VMI chamber to a pressure of 2.2 x 10$^{-7}$ Torr. The voltages applied to the electrostatic lens were measured to be within $\pm$0.2\% of the target voltages shown in Table \ref{tab:electrodes}. Ionization is achieved with a Titanium:Sapphire system producing 30 fs pulses at a repetition rate of 1 kHz. Pulses were spectrally filtered to a 10 nm width centered at 800 nm using an interferometric band pass filter yielding a temporal FWHM (in intensity) of 80 fs with a pulse energy of 12.6 $\mu$J. A wire grid polarizer and $\lambda$/2 plate (both Thorlabs products) are used to clean and orient the laser polarization before being focused into the VMI chamber by a f = +300 mm focusing lens. The intensity in the interaction region is estimated to be on the order of 10$^{13}$ W/cm$^{2}$. The net result of these conditions is that $\sim$15\% of laser pulses have a Xe ionization event.

In Figure \ref{fig:raw} we compare VMI images of Xe$^{+}$ before and after processing. Figure \ref{fig:raw} (a) and (b) present the ion TOF spectrum and clustered VMI image for linearly polarized light oriented along $\hat{y}$. Also included in Figure \ref{fig:raw} (a) is a histogram of cluster TOA. Figure \ref{fig:raw} (b) is clustered, but otherwise there is no data processing, meaning that it includes every pixel activation during data collection. Figure \ref{fig:raw} (c) is a processed image where cluster events are gated by two means. First, we remove clusters with a TOA outside a 1 $\mu$s window (centered about the cluster TOA peak in Figure \ref{fig:raw} (a)) as they are not correlated with electrons emitted by a laser pulse. In doing so $\sim$40\% of clusters are removed, which includes any faulty or dead pixels. Second, we gate on the ion species of interest. Specifically, we only take clusters that occur in coincidence with an ion measured within a particular TOF range. Here we gate on a 1.75 $\mu$s window in the ion TOF (vertical black lines in Figure \ref{fig:raw} (a)) corresponding to Xe$^{+}$.

Much in the same way that the energy resolution for the PCTL-VMI was simulated spatially and temporally, we now require both spatial and temporal energy calibration to access the 3D momentum distribution; this is shown in Figure \ref{fig:calibration}. Spatial calibration is done in a standard manner: ATI peaks for a VMI image (with laser polarization in the xy-plane) are fit under the condition that each is separated by a photon energy and that clusters at R = 0 correspond to having zero kinetic energy. The ATI peaks and corresponding calibration fit are shown to be in good agreement compared to SIMION simulations (Figure \ref{fig:calibration} (a) and (c)). Temporal energy calibration requires a separate dataset to be taken with the laser polarization oriented along $\hat{z}$ so that the ATI are now clearly discernible in the electron TOF spectrum (shown in Figure \ref{fig:calibration} (b)). Notice that in this ATI spectrum there is an additional peak corresponding to zero kinetic energy along $\hat{z}$ present in the electron TOF spectrum. This feature can also be seen in Figure \ref{fig:raw} (c) as the projection of the image on the polarization axis which is now effectively being measured as a TOF distribution. By calibrating the spatial ATI first we have the absolute energy for each remaining (E $\neq$ 0) peak in the electron TOF spectrum. The ATI energies are plotted as a function of TOF and fit to a fifth order polynomial to serve as a conversion from TOF to energy. Figure \ref{fig:calibration} (d) shows the measured ATI TOFs and conversion fit to be in good agreement with SIMION simulations.

\begin{figure}[t!]
\centering
\includegraphics[width=\linewidth]{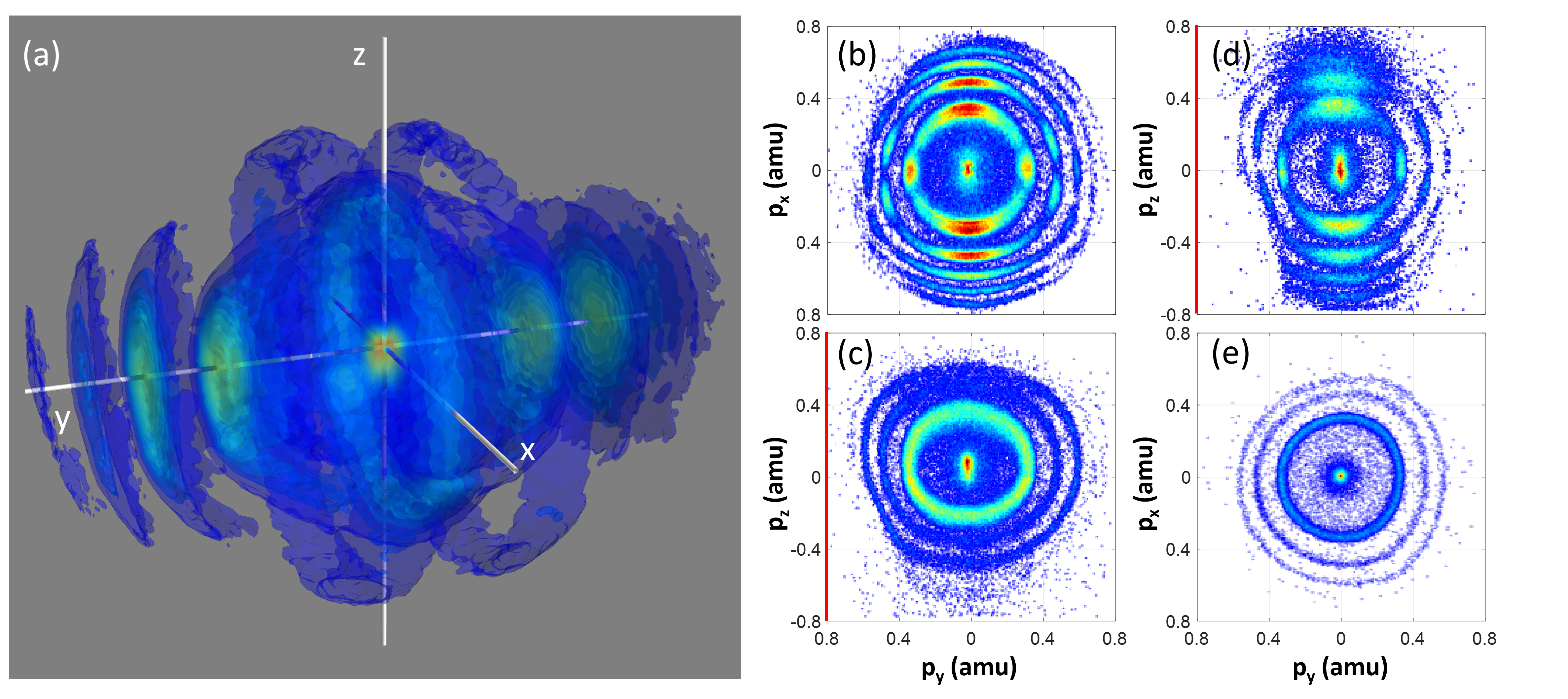}
\caption{\label{fig:threeD} 3D momentum distribution results for singly ionized Xe by linearly polarized light. (a) is a 3D histogram of the electron momentum distribution with the laser polarization along $\hat{y}$ and plotted in atomic momentum units (amu). (b) and (c) are slices from this 3D histogram in the xy and yz plane respectively. (d) and (e) present slices in the yz and xy planes (respectively), but for the laser polarization aligned with the $\hat{z}$-axis. Red lines denote axes for which the momentum is measured by the electron TOF (p$_{z}$).}
\end{figure}

It is important to note that with the PCTL-VMI design the electron TOF spectrum is far broader than with previously reported 3D VMIs. For example, Cheng et al. observe electrons of $\pm$2.12 eV over a range of $\sim$5 ns \cite{cheng20223d}, as compared to $\sim$10 ns for the PCTL-VMI as shown in Figure \ref{fig:calibration}. This means that, for the same timing resolution, the PCTL-VMI design presented here will have greater overall TOF energy resolution. Further, even in the low voltage configuration required for a broad electron TOF spectrum, the PCTL-VMI also retains a higher spatial photoelectron cutoff energy than that shown by Cheng et al.

Figure \ref{fig:threeD} (a) presents the calibrated 3D histogram of the electron momentum distribution, for laser polarization along $\hat{y}$, in coincidence with Xe$^{+}$. This is the same dataset presented in Figure \ref{fig:raw} and Figure \ref{fig:calibration} (a) and (c). Figure \ref{fig:threeD} also presents slices of the 3D histogram for p$_z$ = 0 (b) and p$_x$ = 0 (c) which are then compared to slices from the same experimental conditions, but now with the laser polarized along $\hat{z}$ (Figure \ref{fig:threeD} (d) and (e) anti-respectively). In doing so we have simply rotated the momentum distribution by 90$^{o}$, and, as expected, the p$_z$ = 0 slice for $\hat{y}$ polarized light (b) is in good agreement with the p$_x$ = 0 slice for $\hat{z}$ polarized light (d). The same can be said of Figure \ref{fig:threeD} (c) and (e). Note that this figure shows the VMI lens does exhibit some aberrations, particularly in (c) and (d). From (b) we can see that the momentum distribution is not perfectly centered on the detector in $\hat{y}$, indicating that the laser focus is not properly positioned along this axis. That being the case, we suspect electrons with sufficiently high momentum in -$\hat{z}$ can have trajectories which approach the V1 electrode too closely, altering their course. This results in a measurable effect on the measured y coordinate of the cluster, but is negligible in TOF: an apt description of what is seen in (c) and (d).

\begin{figure*}[t!]
\includegraphics[width=\linewidth]{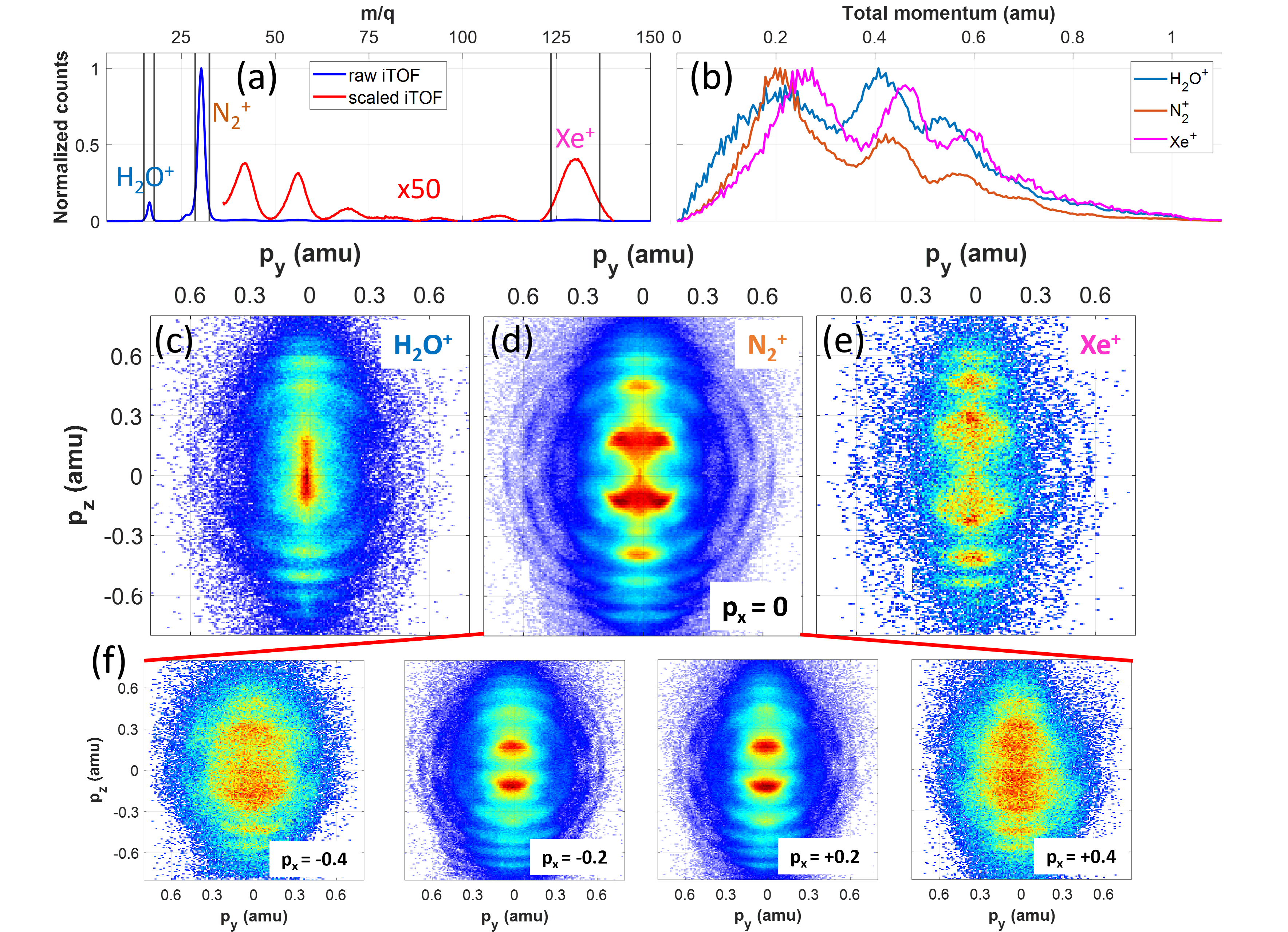}
\caption{\label{fig:coin} Ion coincidence of the 3D electron momentum distribution. (a) is a histogram of ion events in mass to charge ratio (m/q) for a mix of ambient air and a small amount of Xe gas for 240M laser shots. A portion of the histogram is scaled by a factor of 50 and overlain for easy visualization of the less abundant species. The full 3D histogram was gated for electrons measured in coincidence with H$_{2}$O$^{+}$, N$_{2}^{+}$, and Xe$^{+}$ (vertical black lines in (a)) and a histogram of the total (3D) momentum for each is shown in (b). Slices of each 3D histogram at p$_{x}$ = 0 are shown for each ion in (c), (d), and (e). Additionally, slices of N$_{2}^{+}$ for p$_{x}$ = $\pm$0.2 amu, and p$_{x}$ = $\pm$0.4 amu are shown in (f).}
\end{figure*}

From Figure \ref{fig:threeD} we can draw two conclusions. First and foremost, by comparing results for laser polarization along $\hat{y}$ and $\hat{z}$ we can see that the 3D momentum distribution can be accurately retrieved in a single measurement. Particularly, in this we have verified reliable measurement of p$_{z}$ by measurement of TOF. Second, as predicted by the SIMION simulations, the energy (or momentum) resolution along the TOF axis is notably less than that of the spatially resolved axes. For example, splitting of the first ATI ring can be seen in both Figures \ref{fig:raw} (c) and \ref{fig:threeD} (b) which is not observed in \ref{fig:threeD} (d). This can also be seen by comparing the width of each ATI ring in Figure \ref{fig:threeD} (c) and (e). Despite this, having access to the 3D histogram provides information that was previously unavailable from the standard (projected) VMI image, e.g. the structures perpendicular to the polarization axis ($\hat{x}$) in the 2D image Figure \ref{fig:raw} (c) are shown to be rings around $\hat{y}$ in Figure \ref{fig:threeD} (a). Further, each ATI is a spherical shell with preferential emission along the polarization axis. 

Finally, to demonstrate the efficacy of the ion coincidence scheme, a dataset was taken for a mix of ambient air and Xe gas using light polarized in $\hat{z}$. To ionize molecular species in air the pulse energy was raised to 48.8 $\mu$J (as compared to 12.6 $\mu$J for all other results presented). A relatively large dataset was taken, $\sim$240M laser shots over 3 days, to ensure good statistics for each ion species. Figure \ref{fig:coin} (a) shows the ion TOF spectrum for this run; of particular interest are the three dominating species: H$_{2}$O$^{+}$, N$_{2}^{+}$, and Xe$^{+}$. Histograms of the total (3D) momentum were produced for electrons in coincidence with each of these ion species and are shown in Figure \ref{fig:coin} (b). It is seen clearly here that there is a shifting of ATI peaks, due in part to differing ionization potentials (I$_{p}$). This can also be attributed to laser intensity effects, that is, ions with lower I$_{p}$ will be ionized from a larger volume and thus from a larger range of field intensities which results in a change in ATI structure. Further, Figure \ref{fig:coin} (c) through (e) present slices of the 3D histogram for each ion at p$_{x}$ = 0 and (f) presents additional slices for N$_{2}^{+}$ at p$_{x}$ = $\pm$0.2 amu, and p$_{x}$ = $\pm$0.4 amu. This figure shows that not only are the ATI peaks shifted, but there is an entirely different angular momentum structure for each species. The cumulative result is three unique ATI spectra, measured within a single dataset, as expected for three ionic species with vastly different electronic structure.

\section{Conclusion}

We have demonstrated the efficacy of a plano-convex electrostatic field in a thick-lens VMI design to enhance its 3D electron momentum resolution capabilities. SIMION simulations show that, with the simple addition of a mesh electrode to form the plano-convex field, the PCTL-VMI not only retains its high spatial energy resolution but also extends the detectable electron cutoff energy for a given repeller voltage. This is of particular importance for 3D measurements which rely on the electron TOF as they necessitate low repeller voltages to maintain a large enough TOF spread to resolve energetic features. Further, the thick-lens serves to spread the electron TOF distribution, thus requiring less temporal resolution of the electronic detection scheme to attain similar overall energy resolution. We have confirmed these simulations experimentally by demonstrating the PCTL-VMI capability to collect electrons up to $\sim$7 eV with a TOF spread of $\sim$30 ns, both being improvements over previous work by factors of $\sim$1.4 and $\sim$3.75 respectively. Additionally, the PCTL-VMI is equipped with a coincident ion TOF spectrometer which allows for effective gating of electrons from different ionic species in a gas mixture and recovers the unique 3D electron momentum distribution for each. These techniques have the potential to lend themselves to more advanced measurements, particularly involving systems where the electron momentum distributions possess non-trivial symmetries.

\section{Acknowledgments}
The authors would like to recognize the Air Force Office of Scientific Research (AFOSR) for financial support of this work under grant no. FA9550-21-1-0387. T.S. was partially funded by the U.S. Department of Energy (DOE) under grant no. DE-SC0019098. C.C., G.M., and T.W. gratefully acknowledge support from the Department of Energy, Basic Energy Sciences Division, under Award No. DEF-G02-08ER15983.

\section{Data Availability Statement}

The data that support the findings of this study are available from the corresponding author upon reasonable request.

\section{Conflict of Interest}

The authors have no conflicts to disclose.

\bibliographystyle{unsrt}  
\bibliography{bibtex_ref}

\end{document}